\documentclass[twocolumn]{article}
\usepackage{epsfig}
\usepackage{amsmath}

\def\epsilon{\varepsilon}

\title{Shift in the velocity of a front due to a cut-off}
\author{Eric \textsc{Brunet}
    and Bernard \textsc{Derrida} \\
\small Laboratoire de Physique Statistique,
         ENS, 24 rue Lhomond, 75005 Paris, France}
\date{Physical Review E 1997, \textbf{56} (3), 2597--2604}

\begin{document}

\twocolumn[\maketitle\small
\begin{abstract}
   We consider the effect of a small cut-off~$\epsilon$ on the velocity
of a traveling wave in one dimension. Simulations done over more than ten
orders of magnitude as well as a simple theoretical argument indicate
that the effect of the cut-off~$\epsilon$ is to select a single velocity
which converges when~$\epsilon\to0$ to the one predicted by the marginal
stability argument. For small~$\epsilon$, the shift in velocity has the
form~$K(\log\epsilon)^{-2}$ and our prediction for the constant~$K$
agrees very well with the results of our simulations. A very similar
logarithmic shift appears in more complicated situations, in particular in
finite size effects of some microscopic stochastic systems. Our
theoretical approach can also be extended to give a simple way of
deriving the shift in position due to initial conditions in the
Fisher-Kolmogorov or similar equations.
\end{abstract}
\vskip2mm

\par\vskip5mm]

\section{Introduction}

Equations describing the propagation of a front between a stable and an
unstable state appear~\cite{Fisher.Genes.37,KPP.Diffusion.37,%
AronsonWeinberger.Genetics.78,DeeLanger.Propagating.83,%
Bramson.Selection.86,vanSaarloos.MarginalStability.89,%
ColletEckmann.Instabilities.90} in a large variety of situations in
physics, chemistry and biology. One of the simplest equations of this
kind is the Fisher-Kolmogorov~\cite{Fisher.Genes.37,KPP.Diffusion.37}
equation 
\begin{equation}
{\partial h \over \partial t} = {\partial^2h\over\partial x^2} +h-h^3,
\label{eqn:FK}
\end{equation}
which describes the evolution of a space and time dependent concentration
$h(x,t)$ in a reaction-diffusion system. This equation, originally
introduced to study the spread of advantageous genes in a
population~\cite{Fisher.Genes.37}, has been widely used in other contexts, in
particular to describe the time dependence of the concentration of some
species in a chemical reaction~\cite{Kerstein.LatticeGas.86,%
AronsonWeinberger.Genetics.75}.

 For such an equation, the uniform solutions $h=1$ and $h=0$ are
respectively stable and unstable and it is known~\cite{%
AronsonWeinberger.Genetics.78,ColletEckmann.Instabilities.90,%
Bramson.Convergence.83,vanSaarloos.MarginalStability.87,%
vanSaarloos.MarginalStability.88} that for initial conditions such that
$h(x,0) \to 1 $ as $ x \to -\infty$ and $h(x,0) \to 0 $ as $ x \to +
\infty$ there exists a one parameter family $F_v$ of traveling wave
solutions  (indexed by their velocity~$v$) of the form
\begin{equation}
h(x,t)=F_v(x-vt),
\label{eqn:sol}
\end{equation}
with $F_v$ decreasing,  $F_v(z) \to 1$ as $z \to - \infty$ and $F_v(z)
\to 0$ as $z \to \infty$. The analytic expression of the shape $F_v$ is
in general not known but one can determine the range of velocities~$v$ for
which solutions of type (\ref{eqn:sol}) exist.  If one assumes an
exponential decay 
\begin{equation}
F_v(z) \simeq e^{-\gamma z}\qquad\text{for large $z$}, 
\label{eqn:exp}
\end{equation}
it is easy to see by replacing (\ref{eqn:sol}) and (\ref{eqn:exp}) into
(\ref{eqn:FK}) that the velocity $v$ is given by
\begin{equation}
v(\gamma)=\gamma+{1\over\gamma}.
\label{eqn:v(g)1}
\end{equation}

   As $\gamma$ is arbitrary, this shows the well known fact that the
range of possible velocities is $v \geq 2$.  The minimal velocity $v_0=2$
is reached for $\gamma_0=1$ and for steep enough initial conditions
$h(x,0)$ (which decay faster than $e^{-\gamma_0 x}$), the solution
selected~\cite{AronsonWeinberger.Genetics.78,DeeLanger.Propagating.83,%
vanSaarloos.MarginalStability.89,ColletEckmann.Instabilities.90,%
Bramson.Convergence.83,vanSaarloos.MarginalStability.87,%
vanSaarloos.MarginalStability.88} for large~$t$
is the one corresponding to this minimal velocity~$v_0$.

   Equations of type~(\ref{eqn:FK}) are obtained either as the large
scale limit~\cite{Bramson.Selection.86,Kerstein.LatticeGas.86,%
Breuer.FluctuationEffects.94,CookDerrida.Lyapunov.90,%
vanBeijeren.Lyapunov.97,Breuer.MacroscopicLimit.95} or as the mean field
limit~\cite{DerridaSpohn.Polymers.88} of physical situations which are
discrete at the microscopic level (particles, lattice models, etc.) As
the number of particles is an integer, the concentration $h(x,t)$ could
be thought as  being larger than  some $\epsilon$, which would correspond
to the value of $h(x,t)$ when a single particle is present.  Equations of
type (\ref{eqn:FK}) appear then as the limit of the discrete model when
$\epsilon\rightarrow0$. Several authors~\cite{Kerstein.LatticeGas.86,%
Breuer.FluctuationEffects.94,CookDerrida.Lyapunov.90} have already
noticed in their numerical works that the speed $v_\epsilon$ of the
discrete model converges slowly, as $\epsilon$ tends to~0, towards the
minimal velocity $v_0$. We believe that the main effect of having
$\epsilon\neq0$ is to introduce a cut-off in the tail of the front, and
that this changes noticeably the speed.

   The speed of the front is in general governed by its tail. In the
present work, we consider equations similar to (\ref{eqn:FK}), which we
modify in such a way that whenever $h(x,t)$ is much smaller than a
cut-off $\epsilon$, it is replaced by~0. The cut-off~$\epsilon$ can be
introduced by replacing (\ref{eqn:FK}) by
\begin{equation}
{\partial h \over \partial t}={\partial^2h\over\partial
x^2}+(h-h^3)a(h),
\label{eqn:cutoff}
\end{equation}
with
\begin{eqnarray}
\label{eqn:a(h)}
           a(h)&=1    &\qquad\text{if $ h >  \epsilon$},\\
           a(h)&\ll1  &\qquad\text{if $h \ll \epsilon$}.\nonumber
\end{eqnarray}
For example, one could choose  $a(h)=1$ for $h\ge\epsilon$ and
$a(h)=h/\epsilon$ for~$h\le\epsilon$. Another choice that we will use in
section~\ref{sec:calcul} is simply $a(h)=1$ if~$h>\epsilon$ and~$a(h)=0$
if~$h\le\epsilon$.

   The question we address here is the effect of the cut-off $\epsilon$
on the velocity $v_\epsilon$ of the front. We will show that the velocity
$v_\epsilon$ converges, as $\epsilon \to 0$, to the minimal velocity
$v_0$ of the original problem (without cut-off) and that the main
correction to the velocity of the front is
\begin{equation}
v_\epsilon \simeq v_0-{\pi^2\gamma_0^2\over2} \ v''(\gamma_0) \ 
                        {1\over\left(\log\epsilon\right)^2}
\label{eqn:resu}
\end{equation}
for an equation of type~(\ref{eqn:FK}) for which the velocity is
related to the exponential decay $\gamma$ of the shape  (\ref{eqn:sol})
by some relation $v(\gamma)$. (Everywhere we note by  $v_0$ the  minimal
velocity and $\gamma_0$ the corresponding value of the decay $\gamma$.)
In the particular case of equation (\ref{eqn:FK}), where $v(\gamma)$ is
given by~(\ref{eqn:v(g)1}), this becomes 
\begin{equation}
v_\epsilon \simeq 2-{\pi^2\over\left(\log\epsilon\right)^2}.
\label{eqn:resuFK}
\end{equation}

   In section~\ref{sec:model} we describe an equation of
type~(\ref{eqn:FK}) where both space and time are discrete, so that
simulations are much easier to perform. The results of the numerical
simulations of this equation are described in section~\ref{sec:nume}:
as $\epsilon\to0$, the velocity is seen to converge like
$\left(\log\epsilon\right)^{-2}$ to the minimal velocity~$v_0$, and the
shape of the front appears to take a scaling form.

   In section~\ref{sec:calcul} we show that for equations of
type~(\ref{eqn:FK}) in presence of a small cut-off~$\epsilon$ as
in~(\ref{eqn:cutoff}), one can calculate both the shape of the front and
the shift in velocity. The results are in excellent agreement with the
numerical data of section~\ref{sec:nume}.

   In section~\ref{sec:stoch} we consider a model defined, for a finite
number~$N$ of particles, by some microscopic stochastic dynamics which
reduces to the front equation of sections~\ref{sec:nume}
and~\ref{sec:calcul} in the limit $N\to\infty$. Despite the presence of
noise, our simulations indicate that in this case too, the velocity
dependence of the front decays slowly (as $\left(\log N\right)^{-2}$) to
the minimal velocity $v_0$ of the front.

\section{A discrete front equation} \label{sec:model}

   To perform numerical simulations, it is much easier to study a case
where both time and space are discrete variables. We consider here the
equation
\begin{subequations}
\label{eqn:model}
\begin{equation}
     h(x,t+1)= g(x,t)\ \Theta[g(x,t) - \epsilon],
\end{equation}
where
\begin{equation}
 g(x,t)= 1 - \bigl[1 - p h(x-1,t) - (1-p)h(x,t) \bigr]^2.
\end{equation}
\end{subequations}
Time is a discrete variable and if initially the concentration
$h(x,0)$ is only defined when $x$ is an integer, $h(x,t)$ remains so at
any later time. Because $t$ and $x$ are both integers, the cut-off
$\epsilon$ can be introduced as in (\ref{eqn:model}) in the crudest way
using a Heaviside $\Theta$ function. (We have checked however that other
ways of introducing the cut-off $\epsilon$ as in (\ref{eqn:cutoff},
\ref{eqn:a(h)}) do not change the results.)

   Equation (\ref{eqn:model}) appears naturally (in the limit
$\epsilon=0$) in the problem of directed polymers on disordered
trees~\cite{DerridaSpohn.Polymers.88,Derrida.Norvege.91} (where the
energy of the bonds is either~$1$ with probability~$p$ or~$0$ with
probability~$1-p$). At this stage we will not give a justification for
introducing the cut-off~$\epsilon$. This will be discussed in
section~\ref{sec:stoch}.

We consider for the initial condition a step function
\begin{eqnarray}
\label{eqn:init}
  h(x,0)& =0 &\qquad\text{if $x\ge0$,}\\
  h(x,0)& =1 &\qquad\text{if $x<0$.}\nonumber
\end{eqnarray}
Clearly for such an initial condition, $h(x,t)=1$ for  $x < 0$ at all
times. As $h(x,t)\simeq1$ behind the front and $h(x,t)\simeq0 $ ahead of the
front, we define the position $X_t$ of the front at time $t$ by 
\begin{equation}
X_t=\sum_{x=0}^{+\infty} h(x,t).
\label{eqn:pos}
\end{equation}
The velocity of the front $v_\epsilon$ can  then be calculated by
\begin{equation}
v_\epsilon = \lim_{t \to \infty}{ X_t \over t} 
           =\left\langle X_{t+1}-X_t\right\rangle,
\label{eqn:speed}
\end{equation}
where the average is taken over time. (Note that as $h(x,t)$ is only
defined on integers, the difference $X_{t+1}-X_t$ is time dependent and
has to  be averaged as in (\ref{eqn:speed}).)

   When $\epsilon=0$, the evolution equation (\ref{eqn:model}) becomes
\begin{equation}
   h(x,t+1)=  
       1 - \bigl[1 - p h(x-1,t) - (1-p)h(x,t) \bigr]^2.
\label{eqn:mf}
\end{equation}
As for (\ref{eqn:FK}), there is a one parameter family of solutions
$F_v$ of the form (\ref{eqn:sol}) indexed by the velocity $v$ which is
related~(\ref{eqn:exp}) to the exponential decay~$\gamma$ of the shape by
\begin{equation}
v(\gamma)={1\over\gamma}\log\bigl(2pe^{\gamma}+2(1-p)\bigr).
\label{eqn:v(g)2}
\end{equation}
(This relation is obtained as~(\ref{eqn:v(g)1}) by considering the tail of
the front where $h(x,t)$ is small and where therefore (\ref{eqn:mf}) can be
linearized.)

   One can show that for $p<{1\over2}$, $v(\gamma)$ reaches a minimal
value $v_0$ smaller than~1 for some  $\gamma_0$, whereas for
$p\ge{1\over2}$, $v(\gamma)$ is a strictly decreasing function of
$\gamma$, implying that the minimal velocity is $v_0=
\lim\limits_{\gamma\to\infty}v(\gamma)=1$.

   We will not discuss here this phase transition and we assume from now
on that $p<{1\over2}$. Table~\ref{tab:limspeed} gives some values of~$v_0$ 
and~$\gamma_0$ obtained from~(\ref{eqn:v(g)2}).
\begin{table}[h] 
\begin{tabular}{llll}
\hline\hline
$p$        &	0.05	    &	    0.25       & 0.45	 \\
\hline
$\gamma_0$ & 2.751\,111\ldots & 2.553\,244\ldots & 4.051\,851\ldots \\
$v_0$ 	   & 0.451\,818\ldots & 0.810\,710\ldots & 0.979\,187\ldots \\
\hline\hline
\end{tabular}
\caption{Values of $\gamma_0$ and $v_0$ for some $p$ when $\epsilon =0$.}
\label{tab:limspeed}
\end{table}

   It is important to notice that for $p<{1\over2}$, the
function $v(\gamma)$ has a single minimum at $\gamma_0$. Therefore, there
are in general two choices $\gamma_1$ and $\gamma_2$ of $\gamma$ for each
velocity $v$. For $v\neq v_0$, the exponential decay of $F_v(z)$ is dominated by
min$(\gamma_1,\gamma_2)$.  As $v \to v_0$, the two roots $\gamma_1$ and
$\gamma_2$ become equal, and the effect of this degeneracy gives (in
a well chosen frame)
\begin{equation}
F_{v_0}(z) \simeq A\, z\,e^{-\gamma_0z} \qquad \text{for large $z$,}
\label{eqn:general}
\end{equation}
where $A$ is a constant. This large $z$ behavior can be recovered by
looking at the general solution of the linearized form of
equation~(\ref{eqn:mf})
\begin{equation}
   h(x,t+1)=  
       2 p h(x-1,t) +2 (1-p)h(x,t). 
\label{eqn:linear}
\end{equation}

\section{Numerical determination of the velocity}\label{sec:nume}

We iterated numerically~(\ref{eqn:model}) with the initial condition
(\ref{eqn:init}) for several choices of $p < {1\over2}$ and for
$\epsilon$ varying between $0.03$ and $10^{-17}$. We observed that  the
speed is usually very easy to measure because, after a short transient
time, the system reaches a periodic regime for which
\begin{equation}
h(x,t+T)=h(x-Y,t)
\label{eqn:cycle}
\end{equation}
for some constants $T$  and $Y$.  The speed $v_\epsilon$ of the front is
then simply given by
\begin{equation}
v_\epsilon={Y\over T}.
\end{equation}
For example, for $p=0.25$ and~$\epsilon=10^{-5}$, we find $T=431$ and
$Y=343$ so that $v_\epsilon=343/431$. The emergence of this periodic
behavior is due to the locking of the dynamical system of the $h(x,t)$ on
a limit cycle. Because $Y$ and $T$ are integers, our numerical
simulations give the speed with an \emph{infinite accuracy.}

\begin{figure}[ht]
\center{\epsfig{file=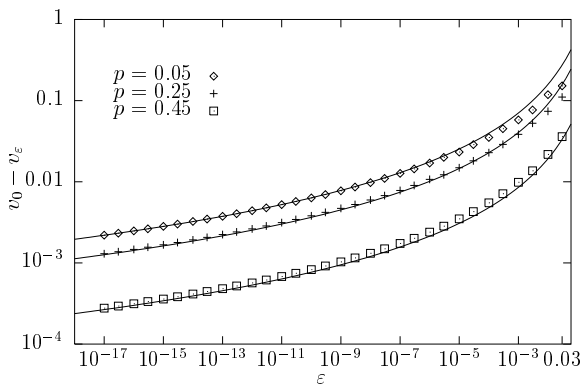, width=0.97\linewidth}}
\caption{The difference $v_0-v_\epsilon$ for $p=0.05$, $0.25$ and $0.5$. The
symbols represent the result of our numerical simulations and the solid lines
indicate  the prediction  of the analysis of section~\ref{sec:calcul}.}
\label{fig:vit}
\end{figure}

   For each choice of $p$ and $\epsilon$, we measured the speed of the
front, as defined by (\ref{eqn:speed}) and its shape.
Figure~\ref{fig:vit} is a log-log plot of the difference $v_0-v_\epsilon$
versus~$\epsilon$ (varying between $0.03$ to~$10^{-17}$) for three
choices of the parameter $p$. The solid lines on the plot indicate the
value predicted by the calculations of section~\ref{sec:calcul}.

   We see on this figure that the velocity $v_\epsilon$ converges
slowly towards the minimal velocity~$v_0$ as $\epsilon\rightarrow0$. Our
simulations, done over several orders of magnitude (here, fifteen),
reveal that the convergence is logarithmic: $v_0 - v_\epsilon \sim
\left(\log\epsilon\right)^{-2}$.

   As the front is moving, to measure its shape, we need to locate its
position. Here we use expression (\ref{eqn:pos}) and we measure the
shape $s_\epsilon(z)$  of the front at a given time~$t$  relative to its
position $X_t$ by
\begin{equation}
s_\epsilon(z)=h(z+X_t,t)
\end{equation}

\begin{figure}[h]
\center{\epsfig{file=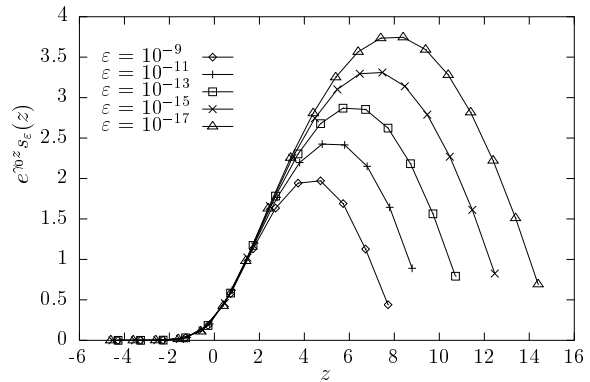, width=0.97\linewidth}}
\caption{Normalized shape of the front $s_\epsilon(z)e^{\gamma_0z}$
versus $z$ for $p=0.25$ and several choices of $\epsilon$.}
\label{fig:shape}
\end{figure}

   When the system reaches the limit cycle (\ref{eqn:cycle}), the shape
$s_\epsilon(z)$ becomes roughly independent of the time chosen.  (In fact
it becomes periodic of period~$T$, but the shape~$s_\epsilon$ has a
smooth envelope.) We have measured this shape at some arbitrary large
enough time to avoid transient effects.  As we expect $s_\epsilon(z)$  to
look more and more like $F_{v_0}(z)$ as $\epsilon$ tends to 0, we
normalize this shape by dividing it by $e^{-\gamma_0z}$. The result
$s_\epsilon(z) e^{\gamma_0z} $ is plotted versus $z$ for $p=0.25$ and
$\epsilon=10^{-9}$, $10^{-11}$, $10^{-13}$, $10^{-15}$ and~$10^{-17}$ in
figure~\ref{fig:shape}.

   On the left part of the graph, our data coincide over an increasing
range as $\epsilon$ decreases, indicating that far from the cut-off, the
shape converges to expression~(\ref{eqn:general}) of
$F_{v_0}(z)$. On the right part, the curves increase up to a maximum
before falling down to some small value which seems to be
independent of~$\epsilon$. When $\epsilon$ is multiplied by a constant
factor (here $10^{-2}$), the maximum as well as the right part of the
curves are translated by a constant amount. This indicates that for
$\epsilon$ small enough, the shape $s_\epsilon(z)$ in the tail (that is
for $z$ large) takes the scaling form
\begin{equation}
s_\epsilon(z) \simeq \left|\log\epsilon\right|\ 
                     G\left({z\over\left|\log \epsilon\right|}\right)\
		     e^{-\gamma_0z}.
\label{eqn:scaling}
\end{equation}

 We will see that our analysis of section~\ref{sec:calcul} does predict
this scaling form. As one expects this shape to coincide with the
asymptotic form~(\ref{eqn:general}) of  $F_{v_0}(z)$ for $ 1 \ll z \ll
|\log \epsilon|$, the scaling function $G(y)$ should be linear for
small~$y$.

\section{Calculation of the velocity for a small cut-off}\label{sec:calcul}

The first remark we make is that as soon as we introduce a cut-off
through a function $a(h)$ which is everywhere smaller than 1, the
velocity~$v_\epsilon$ of the front is lowered compared to the velocity
obtained in the absence of a cut-off.  This is easy to check by comparing
a solution $h_\epsilon(x,t)$  of (\ref{eqn:cutoff}) where $a(h)$ is
present and a solution $h_0(x,t)$  of (\ref{eqn:FK}).  If initially
$h_\epsilon(x,0) < h_0(x,0)$, the solution $h_\epsilon$ will never be
able to take over the solution $h_0$. Indeed, would the two functions
$h_\epsilon(x,0)$ and $h_0(x,0)$ coincide for the first time at some
point $x$, we would have at that point $\partial^2 h_\epsilon / \partial
x^2 \leq \partial^2 h_0 / \partial x^2 $ and together with the effect of
$a(h)$ this would bring back the system in the situation where
$h_\epsilon(x,t) < h_0(x,t)$~\cite{AronsonWeinberger.Genetics.78,%
ColletEckmann.Instabilities.90}. This shows that $v_\epsilon\le v_0$.

For the calculation of the velocity $v_\epsilon$, we will consider
first the modified Fisher-Kolmogorov equation~(\ref{eqn:cutoff})
when the cut-off function $a(h)$ is simply given by
\begin{equation}
  a(h)=  \Theta(h- \epsilon).
\label{eqn:a(h)1}
\end{equation}

In this section we will calculate the leading correction to the velocity
when $\epsilon $ is small and we will obtain the scaling function $G$
which appears in (\ref{eqn:scaling}). Then we will discuss briefly how
our analysis could be extended to more general forms of the cut-off
function $a(h)$ or to other traveling wave equations such as
(\ref{eqn:model}).

As $v_\epsilon$ is the velocity of the front, its shape 
$s_\epsilon(z)=h(z+v_\epsilon t,t)$ in the asymptotic regime satisfies
\begin{displaymath}
 v_\epsilon s_\epsilon' + s_\epsilon'' + (s_\epsilon-s_\epsilon^2)
 a(s_\epsilon) =0.
\end{displaymath}
When $\epsilon$ is small, with the choice (\ref{eqn:a(h)1}) for $a(h)$, we
can decompose the range of values of $z$ into three regions:

\begin{quote}
\begin{description}
\item[Region I]   where $s_\epsilon(z)$ is not small compared to~1.
\item[Region II]  where $\epsilon < s_\epsilon(z) \ll 1$.
\item[Region III] where $s_\epsilon(z) <\epsilon$.
\end{description}
\end{quote}

In region~I, the shape of the front $s_\epsilon$ looks like $F_{v_0}$ 
whereas in regions~II and~III, as $s_\epsilon$ is small, it satisfies
the linear equations
\begin{eqnarray}
v_\epsilon s_\epsilon'+s_\epsilon''+s_\epsilon=0 &\qquad\text{in region~II,}
\label{eqn:linearII} \\
v_\epsilon s_\epsilon'+s_\epsilon''=0            &\qquad\text{in region~III.}
\label{eqn:linearIII} 
\end{eqnarray}
These linear equations (\ref{eqn:linearII},\ref{eqn:linearIII}) can be
solved easily. The only problem is to make sure that the solution in
region~II and its derivative coincides with $F_{v_0}$ at the boundary
between ~I and~II and with the solution valid in region~III at the
boundary between~II and~III. If we call~$\Delta$ the shift in the
velocity 
\begin{equation}
\Delta=v_0-v_\epsilon,
\end{equation} 
and if we note $\gamma_r \pm i \gamma_i$ the two roots of the equation
$v(\gamma) = v_\epsilon$,  the shape $s_\epsilon$ is given in the three
regions by
\begin{eqnarray}
\label{eqn:3ways}
s_\epsilon(z) \simeq & F_{v_0}(z) 
	    &\text{in region I,} \nonumber\\
s_\epsilon(z) \simeq & C e^{-\gamma_r z} \sin (\gamma_i z +D) 
	    &\text{in region II,} \\
s_\epsilon(z) \simeq & \epsilon e^{-v_\epsilon (z-z_0)}
	    &\text{in region III,} \nonumber
\end{eqnarray}
and we can determine the unknown quantities $C$, $D$, $z_0$ and $v_\epsilon$
by using the boundary conditions.

   For large~$z$ we know from~(\ref{eqn:general}) that $F_{v_0}(z) \simeq A z
e^{-\gamma_0 z}$ for some $A$. Therefore, as $\gamma_0- \gamma_r \sim
\Delta$ and $\gamma_i \sim \Delta^{1/2}$, the boundary conditions between
regions~I and~II impose, to leading order in $\Delta^{1/2}$, that $C= A/
\gamma_i$ and $D=0$.

   At the boundary between regions~II and~III, we have
$s_\epsilon(z)=\epsilon$ and $z=z_0$. If we impose the continuity of
$s_\epsilon$ and of its first derivative at this point, we get
\begin{subequations}
\label{eqn:system}
\begin{equation}
      A e^{-\gamma_r z_0} \sin (\gamma_i z_0) = \epsilon\gamma_i,
\end{equation}
and
\begin{equation}
      A e^{-\gamma_r z_0} [ -\gamma_r \sin (\gamma_i z_0) 
               +\gamma_i \cos (\gamma_i z_0)] = - v_\epsilon \epsilon\gamma_i. 
\end{equation}
\end{subequations}
Taking the ratio between these two relations leads to
\begin{equation}
\gamma_r - {\gamma_i \over \tan(\gamma_i z_0)} = v_\epsilon.
\label{eqn:eqq}
\end{equation}

When $\Delta$ is small, $\gamma_r \simeq \gamma_0 = 1$, $v_\epsilon \simeq
v_0 = 2$ and $\gamma_i \sim \Delta^{1/2}$. Thus
the only way to satisfy (\ref{eqn:eqq})  is to set $ \gamma_i z_0 \simeq
\pi$ and $\pi - \gamma_i z_0 \simeq \gamma_i \sim
\Delta^{1/2}$.  Therefore, (\ref{eqn:system}) implies to leading order
that $z_0 \simeq -(\log \epsilon)/ \gamma_0$ and the condition $\gamma_i
z_0 \simeq \pi$ gives
\begin{equation}
\gamma_i \simeq {\pi \over z_0} \simeq {\pi \gamma_0 \over \left|\log
\epsilon\right|} 
\label{eqn:gammai}
\end{equation}
Then, as $\gamma_i$ is small, the difference $\Delta=v_0-v_\epsilon$ is
given by 
\begin{equation}
v_0 - v_\epsilon \simeq {1 \over 2} v''(\gamma_0) \gamma_i^2 
                 \simeq { v''(\gamma_0) \pi^2 \gamma_0^2 \over 
		          2 \left(\log \epsilon\right)^2}
\label{eqn:prediction}
\end{equation}
which is the result announced in~(\ref{eqn:resu}) and~(\ref{eqn:resuFK}).

 A different cut-off function $a(h)$ should not affect the shape of
$s_\epsilon$ in the region~II or the size~$z_0$ of region~II. Only the
precise matching between regions~II and~III might be modified  and we do
not think that this would change the leading dependency of $z_0$ in
$\epsilon$ which controls everything. In fact there are other choices of
the cut-off function $a(h)$ (piecewise constant) for which we could find
the explicit solution in region~III, confirming that the precise
form of $a(h)$ does not change (\ref{eqn:gammai}). The generalization of the
above argument to equations other than~(\ref{eqn:FK}) (and in particular
to the case studied in sections~\ref{sec:model} and~\ref{sec:nume}) is
straightforward. Only the form of the linear equation is changed and the
only effect on the final result~(\ref{eqn:resu}) is that one has to use a
different function~$v(\gamma)$.

   When expression~(\ref{eqn:resu}) is compared in figure~\ref{fig:vit}
to the results  of the simulations, the agreement is excellent.
Moreover, in region~II, one sees from (\ref{eqn:3ways})
and~(\ref{eqn:gammai}) that 
\begin{equation} 
   s_\epsilon(z) \simeq
{A\over\pi\gamma_0}\left|\log\epsilon\right| \sin \left(  \pi \gamma_0 z \over
\left|\log \epsilon \right|\right) e^{-\gamma_0 z},
\label{eqn:scaling2}
\end{equation}
which also agrees with the scaling form (\ref{eqn:scaling}). 

   Recently, for a simple model of evolution~\cite{Kessler.Landscape.96,%
Tsimring.RNA.96} governed by a linear equation, the velocity was found to
be the logarithm of the cut-off to the power~$1\over3$. This result was
obtained by an analysis which has some similarities to the one presented
in this section.

\section{A stochastic model}\label{sec:stoch}

   Many models described by traveling wave equations originate from a
large scale limit of microscopic stochastic models involving a finite
number~$N$ of particles~\cite{Breuer.FluctuationEffects.94,%
CookDerrida.Lyapunov.90,vanBeijeren.Lyapunov.97,%
Breuer.MacroscopicLimit.95}. Here we study such a microscopic model, the
limit of which reduces to~(\ref{eqn:mf}) when~$N\to\infty$.  Our
numerical results, presented below, indicate a large~$N$ correction to
the velocity of the form $v_N\simeq v_0 - a \left(\log N\right)^{-2}$
with a coefficient~$a$ consistent with the one calculated in
section~\ref{sec:calcul} for $\epsilon={1\over N}$.

  The model we consider in this section appears in the study of directed
polymers~\cite{CookDerrida.Lyapunov.90} and is, up to minor changes,
equivalent to a model describing the dynamics of hard
spheres~\cite{vanBeijeren.Lyapunov.97}. It is a stochastic process
discrete both in time and space with two parameters: $N$, the number of
particles, and~$p$, a real number between~0 and~1.   At time~$t$ ($t$ is
an integer), we have $N$ particles on a line at integer positions
$x_1(t)$, $x_2(t)$, \ldots, $x_N(t)$.  Several particles may occupy the
same site. At each time-step, the $N$~positions evolve in the following
way: for each~$i$, we choose two particles~$j_i$ and~$j'_i$ at random
among the $N$~particles. (These two particles do not need to be
different.) Then we update $x_i(t)$ by
\begin{equation}
  x_i(t+1)= \max\bigl(x_{j_i}(t) + \alpha_i,\,
            x_{j'_i}(t) + \alpha'_i\bigr),
\label{eqn:stochmodel}
\end{equation}
where $\alpha_i$ and~$\alpha'_i$ are two independent random numbers taking
the value~1 with probability~$p$ or~0 with probability~$1-p$.  The
numbers $\alpha_i$, $\alpha'_i$, $j_i$ and $j'_i$ change at each
time-step.  Initially~($t=0$), all particles are at the origin so that we
have~$x_i(0)=0$ for all~$i$.

   At time~$t$, the distribution of the $x_i(t)$ on the line can be
represented by a function~$h(x,t)$ which counts the fraction of
particles strictly at the right of~$x$.
\begin{equation}
h(x,t)={1\over N}\sum_{x_i(t)>x} 1.
\end{equation}
Obviously $h(x,t)$ is always an integral multiple of~$1\over N$. At~$t=0$,
we have $h(x,0)=1$ if~$x<0$ and $h(x,0)=0$ if~$x\ge0$.  One can notice
that the definition of the position~$X_t$ of the front used
in~(\ref{eqn:pos}) coincides with the average position of the $N$
particles
\begin{equation}
X_t=\sum_{x=0}^{+\infty} h(x,t) = {1\over N}\sum_{i=1}^N x_i(t).
\end{equation}

   Given the positions $x_i(t)$ of all the particles (or, equivalently¸
given the function $h(x,t)$), the $x_i(t+1)$ become independent random
variables. Therefore, given $h(x,t)$, the probability for each
particle to have at time~$t+1$ a position strictly larger than~$x$ is
given by
\begin{align} 
\label{eqn:stoch}
&\bigl\langle h(x,t+1)\,|\,h(x,t)\bigr\rangle\\
&\qquad = 1-\bigl[1-ph(x-1,t)- (1-p)h(x,t) \bigr]^2.\nonumber
\end{align}

   The difficulty of the problem comes from the fact that one can only
average $h(x,t+1)$ over a single time-step. On the right hand side
of~(\ref{eqn:stoch}) we see terms like $h^2(x,t)$ or $h(x-1,t)h(x,t)$ and
one has to calculate all the correlations of the $h(x,t)$ in order to
find $\bigl\langle h(x,t+1)\bigr\rangle$. This makes the problem very
difficult for finite~$N$. However, given $h(x,t)$, the $x_i(t+1)$ are
independent and in the limit~$N\to\infty$, the fluctuations of~$h(x,t+1)$
are negligible.  Therefore, when~$N\to\infty$, $h(x,t)$ evolves according
to the deterministic equation~(\ref{eqn:mf}).  As the initial condition
is a step function, we expect the front to move, in the
limit~$N\to\infty$, with the minimal velocity~$v_0$ of~(\ref{eqn:v(g)2}).

   For large but finite~$N$, we expect the correction to the velocity to
have two main origins. First, $h(x,t)$ takes only values which are
integral multiples of~$1\over N$, so that $1\over N$ plays a role similar
to the cut-off $\epsilon$ of section~\ref{sec:model}.   Second, $h(x,t)$
fluctuates around its average and this has the effect of adding noise to
the evolution equation~(\ref{eqn:mf}). In the rest of this section we
present the results of simulations done for large but finite~$N$ and we
will see that the shift in the velocity seems to be very close to the
expression of section~\ref{sec:calcul} when~$\epsilon={1\over N}$.

   With the most direct way of simulating the model for $N$ finite, it is
difficult to study systems of size much larger than~$10^6$. Here we use a
more sophisticated method allowing $N$ to become huge. Our method, which
handles many particles at the same time, consists in iterating directly
$h(x,t)$.

\def\xmin{x_{\text{min}}}\def\xmax{x_{\text{max}}} 

   Knowing the function~$h(x,t)$ at time~$t$, we want to
calculate~$h(x,t+1)$. We call respectively $\xmin$ and~$\xmax$ the
positions of the leftmost and rightmost particles at time~$t$
and~$l=\xmax-\xmin+1$. In terms of the function~$h(x,t)$, one has
$0<h(x,t)<1$ if and only if $\xmin\le x < \xmax$.
Obviously, all the positions~$x_i(t+1)$ will lie
between~$\xmin$ and~$\xmax+1$. The probability~$p_k$ that a given
particle~$i$ will be located at position~$\xmin+k$ at time~$t+1$ is
\begin{equation}
  p_k=\bigl\langle h(\xmin+k-1,t+1)\bigr\rangle-\bigl\langle
  h(\xmin+k,t+1)\bigr\rangle,
\end{equation}
with $\bigl\langle h(x,t+1)\bigr\rangle$ given
by~(\ref{eqn:stoch}). Obviously, $p_k\neq0$ only for $0 \le k \le l$.

   The probability to have, for every $k$, $n_k$ particles at location
 $\xmin+k$ at time~$t+1$ is given by
\begin{align}
P(n_0,n_1,\ldots,n_l)=&{N!\over n_0!\,n_1!\,\ldots\,n_l!}\
      p_0^{n_0}\,p_1^{n_1}\,\ldots\,p_l^{n_l}\nonumber\\
      &\times\delta(N-n_0-n_1-\cdots-n_l).
\label{eqn:proba}
\end{align}

   Using a random number generator for a binomial distribution,
expression~(\ref{eqn:proba}) allows to generate random~$n_k$.  This is
done by calculating $n_0$ according to the distribution 
\begin{equation}
P(n_0)={N!\over n_0!\,(N-n_0)!}\ p_0^{n_0}(1-p_0)^{N-n_0},
\end{equation}
then $n_1$ with
\begin{eqnarray}
P(n_1 \,|\, n_0)
            &=& {(N-n_0)!\over n_1!\,(N-n_0-n_1)!}\
         \left(p_1\over 1-p_0\right)^{n_1}\nonumber\\
	    & &\times 
         \left(1-{p_1\over 1-p_0}\right)^{N-n_0-n_1},
\end{eqnarray}
and so on. This method can be iterated to produce the $l+1$~numbers
$n_0$, $n_1$,~\ldots, $n_l$ distributed according to~(\ref{eqn:proba}).
Then we construct~$h(x,t+1)$ by 
\begin{align}
h(x,t+1)=& 1 && \text{if $x<\xmin$,}\nonumber\\
h(x,t+1)=& \displaystyle{1\over N} \sum_{i=k+1}^l n_i 
     		&& \vcenter{\hbox{if $\xmin\le x\le\xmax+1$}
	                     \hbox{and $x=\xmin+k$,}}\nonumber\\
h(x,t+1)=&  0	&& \text{if $x>\xmax+1$.}
\end{align}
As the width~$l$ of the front is roughly of order $\log N$, this method allows
$N$ to be very large.

\begin{figure}[ht]
\center{\epsfig{file=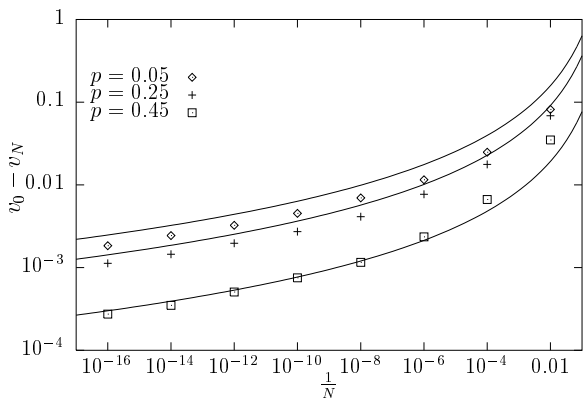, width=0.97\linewidth}}
\caption{The difference $v_0-v_N$ versus $1\over N$ for three
choices of $p$. The symbols represent the result of our numerical simulations
of the stochastic process and the solid lines indicate the
prediction~(\ref{eqn:resu}) for $\epsilon={1\over N}$.}
\label{fig:vitR}
\end{figure}

   Using this method with the generator of random binomial numbers
given in~\cite{NumericalRecipes}, we have measured the velocity~$v_N$ 
of the front for
several choices of~$p$ (0.05, 0.25 and~0.45) and for $N$ ranging from~100
to~$10^{16}$. We measured the velocities with the expression
\begin{equation}
v_N={X_{10^6}-X_{10^5} \over 9\,10^5}.
\end{equation}   
Figure~\ref{fig:vitR} is a log-log plot of the difference $v_0-v_N$
versus $1\over N$ compared to the prediction~(\ref{eqn:resu})
for $\epsilon={1\over N}$. The variation of $v_N$ when using longer times
or different random numbers were not larger than the size of the symbols.

   We see on figure~\ref{fig:vitR} that the speed~$v_N$ of the front seems to be
given for large~$N$ by
\begin{equation}
v_N\simeq v_0 - {K\over \left(\log N\right)^2},
\end{equation}
where the coefficient $K$ is not too different from the
prediction~(\ref{eqn:resu}).

   The agreement is however not perfect. The shift~$v_0-v_N$ seems to be
proportional to~$\left(\log N\right)^{-2}$, but the constant looks on
figure~\ref{fig:vitR} slightly different from the one predicted
by~(\ref{eqn:resu}). A possible reason for this difference could have
been the discretization of the front: instead of only cutting off the
tail as in sections~\ref{sec:nume} and~\ref{sec:calcul}, here the whole
front~$h(x,t)$ is constrained to take values multiple of $1\over N$. One
might think that this could explain this discrepancy.  However, we have
checked numerically (the results are not presented in this paper) that
equation~(\ref{eqn:mf}) with $h(x,t)$ constrained to be a multiple of a
cut-off $\epsilon$ does not give results significantly different from the
simpler model of sections~\ref{sec:nume} and~\ref{sec:calcul}
with only a single cut-off. So we think that the full
discretization of the front can not be responsible for a different
constant~$K$. The discrepancy observed in figure~\ref{fig:vitR} is more
likely due to the effect of the randomness of the process. It is however
not clear whether this mismatch would decrease for even larger~$N$. It
would be interesting to push further the numerical simulations and check
the $N$-dependence of the front velocity for very large $N$.

\section{Conclusion}

   We have shown in the present  work that a small cut-off $\epsilon$ in
the tail of solutions of traveling wave equations has the effect of
selecting a single velocity~$v_\epsilon$ for the front. This
velocity~$v_\epsilon$ converges to the minimal velocity~$v_0$ when
$\epsilon\to0$ and the shift~$v_0-v_\epsilon$ is surprisingly
large~(\ref{eqn:resu},~\ref{eqn:resuFK}).

   Very slow convergences to the minimal velocity have been observed in a
number of cases~\cite{Kerstein.LatticeGas.86,%
Breuer.FluctuationEffects.94,CookDerrida.Lyapunov.90,%
vanBeijeren.Lyapunov.97} as well as the example of
section~\ref{sec:stoch}. As the effect of the cut-off on the velocity is
large, it is reasonable to think that it would not be much affected by
the presence of noise. The example of section~\ref{sec:stoch} shows that
the cut-off alone gives at least the right order of magnitude for the
shift and it would certainly be interesting to push further the
simulations for this particular model to see whether the analysis
of section~\ref{sec:calcul} should be modified by the noise. The numerical
method used in section~\ref{sec:stoch} to study a very large ($N\sim
10^{16}$) system was very helpful to observe a logarithmic behavior. We
did not succeed to check in earlier
works~\cite{Breuer.FluctuationEffects.94,CookDerrida.Lyapunov.90,%
vanBeijeren.Lyapunov.97,Kerstein.TwoParticle.88} whether the correction 
was logarithmic, mostly because the published data were usually too noisy
or obtained on a too small range of the parameters.  Still even if the
cut-off was giving the main contribution to the shift of the velocity,
other properties would remain very specific  to the presence of  noise
like the diffusion of the position of the
front~\cite{Breuer.MacroscopicLimit.95}.

Our approach of section~\ref{sec:calcul} shows that the effect of a small
cut-off is the existence of a scaling
form~(\ref{eqn:scaling},~\ref{eqn:scaling2}) which describes the change
in the shape of the front in its steady state. The effect of initial
conditions for usual traveling wave equations (with no cut-off) leads to
a very similar scaling form for the change in the shape of the front in
the transient regime. This is explained in the appendix where
we show how the logarithmic shift of the position of a front due to
initial conditions~\cite{Bramson.Convergence.83,Bramson.Displacement.78}
can be recovered.

\setcounter{section}{0}\def\thesection{\Alph{section}}  
\numberwithin{equation}{section}

\section{Effect of initial conditions on the position and on the shape of
	 the front}
\label{app:init}

   In this appendix we show that ideas very similar to those developed in
section~\ref{sec:calcul} allow one to calculate the position and the shape at
time~$t$ of a front evolving according to~(\ref{eqn:FK}), or to a similar
equation, given its initial shape. The main idea is that in the long time
limit, there is a region of size $\sqrt{t}$ ahead of the front which
keeps the memory of the initial condition. We will recover in particular
the logarithmic shift in the position of the front due to the initial
condition~\cite{Bramson.Convergence.83,Bramson.Displacement.78}, namely
that if the initial shape is a step function
\begin{eqnarray}
h(x,0)=& 0 & \qquad\text{if $x>0$,}\\
h(x,0)=& 1 & \qquad\text{if $x<0$,}\nonumber
\end{eqnarray}
then the position~$X_t$ of the front at time~$t$ increases like
\begin{equation}
  X_t\simeq 2t-{3\over2} \log t.
\label{eqn:pos3/2}
\end{equation}
More generally, if initially
\begin{eqnarray}
  h(x,0)=& x^\nu e^{-\gamma_0x} & \qquad\text{if $x>0$,}\\
  h(x,0)=& 1                    & \qquad\text{if $x<0$,}\nonumber
\end{eqnarray}
we will show that for $\nu>-2$
\begin{equation}
X_t\simeq 2t+ {\nu-1\over2}\log t,
\label{eqn:pos(nu-1)/2}
\end{equation}
whereas the shift is given by~(\ref{eqn:pos3/2}) for $\nu<-2$.
Here, there is no cut-off but the transient behavior in the long time
limit gives rise to a scaling function very similar to the one discussed
in section~\ref{sec:calcul}.

If we write the position of the front at time~$t$ as
\begin{equation}
X_t=v_0 t - c(t),
\label{eqn:posb}
\end{equation}
we observed numerically (as in figure~\ref{fig:shape} of 
section~\ref{sec:nume}) and we are going to see
in the following that the shape of the front takes for large~$t$ the
scaling form
\begin{equation}
  h(x,t)=t^\alpha G\left({x-X_t \over t^\alpha}\right)
             e^{-\gamma_0(x-X_t)},
\label{eqn:scalingF}
\end{equation}
very similar to~(\ref{eqn:scaling},~\ref{eqn:scaling2}).

If we use~(\ref{eqn:posb}) and~(\ref{eqn:scalingF}) into the linearized
form of equation~(\ref{eqn:FK}), we get using, the fact that $v_0=2$ and
$\gamma_0=1$,
\begin{equation}
{1\over t^\alpha}G''+{1\over t^{1-\alpha}}\left(\alpha z G'-\alpha
G\right)+t^\alpha \dot{c}\, G= \dot{c}\, G',
\label{eqn:diff1}
\end{equation}
where $z=(x-X_t)t^{-\alpha}$.
By writing that the leading orders of the different terms
of~(\ref{eqn:diff1}) are comparable, we see that we must have
\begin{eqnarray}
  \alpha&=&{1\over2}, \\
  \dot{c}   &\simeq&{\beta\over t}, 
\end{eqnarray}
for some $\beta$, and that the right hand side of~(\ref{eqn:diff1}) is
negligible. Therefore, the equation satisfied by $G$ is
\begin{equation}
{d^2\over dz^2}G+{z\over2} {d\over dz}G +\left(\beta-{1\over2}\right) G =0,
\label{eqn:diff2}
\end{equation}
and the position of the front is given by
\begin{equation}
X_t\simeq v_0t-\beta \log t.
\label{eqn:pos1}
\end{equation}

As in section~\ref{sec:calcul}, we expect that as $t\to\infty$, the front
will approach its limiting form and therefore that for $z$~small, the
shape will look like~(\ref{eqn:general}). Therefore we choose the
solution~$G_\beta(z)$ of~(\ref{eqn:diff2}) which is linear at~$z=0$.
This solution can be written as an infinite sum
\begin{align}
G_\beta(z) &= A\sum_{n=0}^\infty
{(-1)^n\over(2n+1)!}z^{2n+1}\prod_{i=0}^{n-1}(\beta+i) ,\nonumber\\
&=A\sum_{n=0}^\infty
{(-1)^n\over(2n+1)!}\,{\Gamma(n+\beta)\over\Gamma(\beta)}\,z^{2n+1}.
\label{eqn:sum}
\end{align}
(The second expression is not valid when~$\beta$ is a non-positive
integer.)

To determine $\beta$, one can notice that the scaling
form~(\ref{eqn:scalingF}) has to match the initial condition when $x$ is
large and $t$ of order 1. We thus need to calculate the asymptotic
behavior of $G(z)$ when $z$ is large.

For certain values of $\beta$, there exist closed expressions of the
sum~(\ref{eqn:sum}). For instance,
\begin{eqnarray}
\label{eqn:Gbeta}
G_{-2}(z) &=& A\left(z+{z^3\over3}+{z^5\over60}\right),\nonumber\\ 
G_{7\over2}(z) &=&
A\left(z-{z^3\over3}+{z^5\over60}\right)e^{-{z^2\over4}},\nonumber \\
G_{-1}(z) &=& A\left(z+{z^3\over6}\right), \nonumber\\
G_{5\over2}(z)&=& A\left(z-{z^3\over6}\right)e^{-{z^2\over4}}, \\
G_{0}(z)&=& Az, \nonumber\\
G_{3\over2}(z)&=& Aze^{-{z^2\over4}},\nonumber\\
G_1(z) &=& Ae^{-{z^2\over4}}\int_0^z e^{t^2\over4}dt,\nonumber\\
G_{1\over2}(z)&=&A\int_0^ze^{-{t^2\over4}}dt,\nonumber
\end{eqnarray}
One can check directly on~(\ref{eqn:diff2}) that $G_\beta$ has a
symmetry
\begin{equation}
G_\beta(z)=-i e^{-z^2/4}\,G_{{3\over2}-\beta}(iz).
\end{equation}

For any $\beta$, one can obtain the large~$z$
behavior of $G(z)$. To do so, we note that for $\beta>0$, one can
rewrite~(\ref{eqn:sum}) as
\begin{eqnarray}
G_\beta(z)&=&{A\over\Gamma(\beta)}\int_0^\infty dt\; t^{\beta-{3\over2}}
\sin(\sqrt{t} z) e^{-t},\nonumber\\
   &=&{2A\over\Gamma(\beta)}z^{1-2\beta}\int_0^\infty
dt\;t^{2\beta-2} \sin(t) e^{-{t^2\over z^2}}.
\label{eqn:integ}
\end{eqnarray}
For $0<\beta<1$, the second integral in~(\ref{eqn:integ}) has a non
zero limit and this gives the asymptotic behavior of $G_\beta(z)$
\begin{equation}
G_\beta(z)\simeq-{2A\over\Gamma(\beta)}\cos(\pi\beta)\,
\Gamma(2\beta-1) z^{1-2\beta}.
\label{eqn:asymptot}
\end{equation}
From~(\ref{eqn:sum}), one can also show that
\begin{equation}
G_\beta''=-{\Gamma(\beta+1)\over\Gamma(\beta)}G_{\beta+1},
\end{equation}
implying that~(\ref{eqn:asymptot}) remains valid for
all~$\beta$ except for $\beta={3\over2}$, $5\over2$, $7\over2$, etc.,
where the amplitude in~(\ref{eqn:asymptot}) vanishes. 
For these values of~$\beta$, $G_\beta(z)$
decreases faster than a power law (see~(\ref{eqn:Gbeta})).

The functions $G_\beta$ calculated so far are acceptable scaling
functions for the shape of the front only for $\beta\le{3\over2}$. Indeed,
one can see in~(\ref{eqn:asymptot}) that for ${3\over2}<\beta<{5\over2}$
the function $G_\beta(z)$ is negative for large~$z$. In fact, for all
$\beta>{3\over2}$, this function changes its sign at least once, so that
the scaling form~(\ref{eqn:scalingF}) is not reachable for an initial
$h(x,0)$ which is always positive. It is only for $\beta\le{3\over2}$ that
$G_\beta$ remains positive for all~$z>0$.

Looking at the asymptotic form~(\ref{eqn:asymptot}), we see that if
initially $h(x,0)=x^\nu e^{-\gamma_0 x}$, the only function $G_\beta(z)$
which has the right large~$z$ behavior is such that $1-2\beta=\nu$, and
this gives, together with~(\ref{eqn:pos1}), the
expression~(\ref{eqn:pos(nu-1)/2}) for the shift of the position.  As the
cases~$\beta>{3\over2}$ are not reachable, all initial conditions
corresponding to~$\nu<-2$ or steeper (such as step functions) give rise
to $G_{3\over2}$ and the shift in position given by~(\ref{eqn:pos3/2}).

All the analysis of this appendix can be extended to other
traveling wave equations such as~(\ref{eqn:mf}), with more general
functions~$v(\gamma)$ (having a non-degenerate minimum at~$\gamma_0$) as
in~(\ref{eqn:v(g)2}). Then the
expressions~(\ref{eqn:pos3/2},~\ref{eqn:pos(nu-1)/2}) of the shift
become
\begin{equation}
X_t\simeq v_0 t -{3\over2\gamma_0}\log t
\end{equation}
and
\begin{equation}
X_t\simeq v_0 t -{1-\nu\over2\gamma_0}\log t. 
\end{equation}

\bigskip\bigskip

We thank C.~Appert, V.~Hakim and~J.L.~Lebowitz for useful discussions.

\end{document}